\documentclass[aps,titlepage,12pt]{revtex4}
\usepackage{amsfonts}
\usepackage{amsmath}
\usepackage{graphicx}
\usepackage{dcolumn}
\usepackage{bm}
\usepackage{overpic}
\usepackage{booktabs}
\usepackage{color}
\usepackage[justification=raggedright,font={small,sf}, singlelinecheck=false]{caption}
\usepackage{natbib}

\begin{document}
	\title{Estimating the Containment Effectiveness and Economic Cost of Inner-city Non-Pharmaceutical Interventions}
	
	\author{Xihan Zhang$^{1,2}$, Yuqing Liu$^{3,2}$, Chen Zhao$^{2}$\footnote{Email:tczxz007@hebtu.edu.cn}, Guijun Li$^{2}$\footnote{Email: ligj@hebtu.edu.cn}}
	
	\affiliation{$^{1}$College of Computer and Cyber Security, Hebei Normal University, 050024 Shijiazhuang, Hebei, P.R. China\\
		$^{2}$School of Mathematical Sciences, Hebei Normal University, 050024 Shijiazhuang, Hebei, P.R. China\\
		$^{3}$School of Geographical Sciences, Hebei Normal University, 050024 Shijiazhuang, Hebei, P.R. China}

	\begin{abstract}
		Non-pharmaceutical interventions (NPIs) are crucial for controlling pandemics, but existing research often overlooks the heterogeneity of individual behavior, which can lead to inaccurate evaluations of the effectiveness of strategies. In this paper, we use a large dataset of fine-grained real-world individual trajectory data from a major Chinese city to examine the trade-off between the epidemic containment effectiveness and economic cost of different NPIs. Our findings reveal significant variations in the outcomes of different NPIs across activation mechanisms and initial scales of undetected transmission. Based on these results, we construct a two-dimensional evaluation framework that comprehensively evaluates the impact of both the containment effectiveness and economic cost, which suggests that implementing stringent strategies—such as lockdown or contact tracing—at low activation thresholds can achieve optimal epidemic control with minimal economic cost. Our study provides a data-driven decision-making framework for understanding the implementation effectiveness and applicability of emergency management policies within urban systems.
	\end{abstract}
	
	\maketitle
	
	\section{Introduction}
	As is well known, the world experienced a severe pandemic in 2019. The novel coronavirus pneumonia (also known as COVID-19) has rapidly evolved into a public health emergency. According to the World Health Organization (WHO), the global total of confirmed cases had exceeded 770 million, with nearly 7 million deaths reported~\cite{who2023} . The repercussions of the pandemic on the global economy have been profound, leading to the world economy contracting by 3.5\% in 2020, which was the most severe economic downturn since the Great Depression~\cite{imf2021weo,world2022global}. Pandemic-induced disruptions have profoundly altered the global economic landscape, triggering cascading effects across supply chains~\cite{xu2020impacts,ozdemir2022supply}, the service sector~\cite{kim2021uncertainty}, and labor market~\cite{fana2020employment}.
	
	To rapidly curb the spread of the epidemic and protect public health, non-pharmaceutical interventions (NPIs) play a crucial role in achieving short-term, rapid control of epidemic spreading~\cite{hsiang2020effect,flaxman2020estimating,kraemer2020effect}, in addition to pharmaceutical measures such as vaccination~\cite{castioni2024rebound} and the development of antiviral drugs~\cite{beigel2020remdesivir}. The NPIs include wearing masks~\cite{yang2024reconciling}, regional lockdowns~\cite{eyawo2021lockdowns,lau2020positive}, restrictions on gatherings~\cite{prem2020effect}, conducting mass testing~\cite{mercer2021testing}, remote work~\cite{yeung2024telecommuting,mitchell2025care}, and implementing travel restrictions~\cite{gossling2020pandemics}. They aim to alleviate pressure on healthcare systems~\cite{cai2022modeling} by reducing the risk of epidemic spreading, thereby safeguarding public health and public health security.
	
	In the domain of practical public health management, the core mechanism of the majority of NPIs is the regulation of individuals' daily behavioral patterns with a view to reducing exposure risks. However, in the existing epidemiological models, the mechanism underlying the impact of various NPIs on individual behavior is often oversimplified. Within the framework of traditional population-based models (e.g., SIR), which assume individual-level homogeneity, NPIs are inherently conceptualized as modifications to population-level parameters~\cite{flaxman2020estimating,prem2020effect,ge2023effects,liu2021impact}. The traditional models, while useful for depicting macro-level epidemic trends, fail to capture micro-level infection processes due to their neglect of individual contact heterogeneity, which limits their reliability in evaluating the effectiveness of NPIs. In order to incorporate heterogeneity in social contact structures, network-based transmission models have been widely adopted. This type of model constructs a social contact network among individuals by abstracting people as nodes and social relationships as connecting edges. Through the above, the effect of NPIs is no longer represented as a macro-level parameter but is instead mapped as direct manipulations of the network topology, e.g., removing edges between nodes to describe the closure of specific venues~\cite{alrasheed2020covid,aleta2020modelling}. However, these models typically rely on static or predefined network structures, which hinders the capacity to adequately simulate the dynamic heterogeneity observed during individual movement processes. By contrast, agent-based models (ABMs) offer greater flexibility by simulating individual-level dynamics and allowing for time-varying interactions~\cite{yang2024reconciling,bouchnita2020hybrid,atias2025optimal,moreno2025critical}. However, most ABMs incorporate NPIs by adjusting predefined interaction rules that are not grounded in observed behavior, which lack the ability to reflect the heterogeneity and adaptive nature of real-world human responses to policy interventions. Some of the recent research has sought to address these limitations by incorporating empirical mobility data into model calibration~\cite{pangallo2024unequal,wang2023quantifying,kerkmann2025agent}. While such data can inform more realistic movement patterns, they are often synthetic, aggregated, or insufficiently granular to capture fine-scale spatiotemporal contact dynamics. Although a few studies use real-world individual trajectory data to simulate epidemic spreading progression ~\cite{yan2024spatiotemporal,zhao2023high}, leaving the quantitative evaluation of NPIs’ effectiveness largely still unexplored.
	
	The essence of NPIs lies in restricting and modifying the individual's temporal trajectory. Therefore, in order to accurately characterize the effects of NPIs, we should not overlook the heterogeneity of individual mobility. Our study leverages empirical human trajectory data derived from 4G communication records between base stations and mobile phones, including around 3 million users out of a total population of 10 million in the city. The data were provided by one of the three major service providers in Shijiazhuang, the capital city of Hebei Province, China. To identify meaningful locations in the user trajectories, we divide the whole period covered by the dataset into 15-minute intervals (see Materials and Methods). If an individual stayed in the same location for more than 10 minutes in each interval, the location was recorded as a stop for that user. Based on this large-scale, high-resolution, real individual trajectory data, we establish a framework to evaluate the city-level containment effectiveness of NPIs. The above data enable us to precisely identify individual spatiotemporal contacts and interactions, thereby constructing a more realistic transmission model. To fully preserve individual heterogeneity, we apply the NPI mechanisms directly to individual trajectories when simulating containment outcomes. Subsequently, by feeding these containment outcomes into an integrated economic model that captures GDP, healthcare expenditure, and direct intervention cost, we further evaluate the economic implications of NPIs. Notwithstanding a growing body of literature on the economic consequences of NPIs~\cite{pangallo2024unequal,henry2025agent,haw2022optimizing,pichler2022forecasting,ash2022disease,dobson2023balancing}, the effectiveness of such interventions is typically evaluated based on the synthetic outcomes derived from model simulations rather than validated against the analysis by real-world individual mobility data. As discussed, a fundamental limitation of these approaches lies in their inability to capture the heterogeneity and dynamic nature of real-world human behavior, which consequently compromises the reliability of their economic evaluations. By grounding our economic analysis in more accurate containment simulations, we provide a more robust evaluation of the trade-offs between epidemic containment and economic cost. Under different activation mechanisms and initial scales of undetected transmission, the containment effectiveness and economic cost of NPIs vary significantly. The results indicate that as the activation threshold increases, leading to a larger scale of infections, the secondary resurgence of the epidemic caused by certain strategies becomes more pronounced, which in turn leads to a more significant escalation of economic losses. Furthermore, as the initial scale of undetected transmission expands, the phenomenon of healthcare system overload occurs earlier, and economic losses are significantly brought forward.

	To systematically evaluate the effectiveness of each NPI under different epidemic scenarios, we establish a two-dimensional evaluation framework centered on epidemic containment effectiveness and economic cost as core metrics. Among all the NPIs, strict city-wide lockdown and complete isolation by contact tracing prove most effective under low activation thresholds, achieving an optimal balance between epidemic containment and economic cost. The simulation results can therefore provide decision support for policymakers regarding “which NPI to implement under what circumstances”. Furthermore, the analytical framework and methodology we established demonstrate strong universality, in that they are not only applicable to COVID-19 but can also be extended to research on prevention and control strategies for other respiratory infectious diseases by adjusting model parameters. Our study offers significant theoretical guidance and practical implications for improving public health emergency management systems.
	
	\section{Results}
	The real-world mobile trajectory big data used in this study is based on 4G communication records between base stations and mobile phones provided by one of the three major service operators in Shijiazhuang, a city in northern China, during the period from 22 to 28 May 2017. The original data contain more than 3 million users, representing approximately 28\% of the city's total population of 11 million. To identify locations in user trajectories, we divide the entire time period covered by the dataset into 15-minute time windows. If a user remains at a specific location for 10 minutes or longer, continuously or discontinuously, within this time window, that location will be recorded as the user's stopping point. A co-location visit event is defined as occurring when two users visit the same location within the same time window. Additionally, to obtain a valid mobile trajectory dataset, we rigorously filter the raw data (see Materials and Methods), ultimately selecting 702,477 valid individual mobile trajectories. Given the cyclical nature of individuals' daily behavioral patterns, we concatenate the dataset five times to ultimately form a 35-day mobile trajectory dataset for subsequent research.
	
	Fig. 1 illustrates the framework for evaluating NPIs, which encompasses two core components: epidemic control and economic cost. At the epidemic level, we first construct an individual-level transmission model based on co-location visit events. In our model, we incorporate individual attributes, including age, residence, and workplace to capture heterogeneity in health state transition probabilities and contact patterns across different location types (Fig. 1A; see Materials and Methods). Disease transmission is set to occur within and between three types of locations: (1) residence, (2) workplace, and (3) community (Fig. 1B). We focus on the NPIs depicted in Fig. 1D. These interventions are applied to the real-world individual movement trajectories shown in Fig. 1C. Combined with the transmission model, our study can characterize the detailed infection processes for all individuals under NPIs.
	
	At the economic level, to analyze the impact of NPIs on the urban economy, we assign each individual to an industry (Fig. 1A; see Materials and Methods). Based on the transmission model simulations, we track the health status of every individual across all industries at each time step, which enables us to jointly analyze three key economic indicators during the pandemic (see Materials and Methods): GDP, government healthcare expenditure for patients, and the direct cost of implementing NPIs. Together, these indicators provide a comprehensive evaluation of the economic benefits of epidemic prevention and control strategies (Fig. 1E).
	
	\textbf{Baseline Scenario.} Given the sudden emergence of large-scale epidemics and the inherent delays in public health responses, we first simulate the pre-detection phase of an outbreak in the absence of interventions. By the time the outbreak is detected, the transmission model shows that the cumulative number of naturally occurring positive cases has already reached 0.5\% of the urban population (a moderate initial scale of undetected transmission). Based on this initial condition, we simulate the baseline scenario over 35 days without implementing any interventions, which serves as a benchmark for evaluating intervention effectiveness. 
	
	In the absence of government intervention, new infections peak within the first week, triggering a citywide outbreak (Fig. 2A, black line). Notably, this pattern of explosive infections closely mirrors the epidemic trajectory observed in Shijiazhuang in 2022, the first Chinese city to lift all COVID-19 restrictions. This finding validates the reliability of our data-driven transmission model.
	
	At the economic level, the most severe contraction occurs during the second week of the outbreak (Fig. 2D, black dashed line). During this period, a large number of infected individuals are unable to participate in socioeconomic activities, while government healthcare expenditure surged. Calculation shows that average weekly economic output falls by 105\% compared to pre-pandemic levels, indicating a severe recessionary shock to the entire economy. This highlights the need for timely intervention to prevent prolonged and substantial economic damage.
	
	\textbf{Outcomes of NPIs Under Different Activation Mechanisms.} We select five NPIs primarily implemented in epidemic prevention and control, namely: (1) the 1 km activity radius restriction; (2) lockdown; (3) isolation; (4) contact tracing; and (5) the closure of high-risk industries (see details in Materials and Methods). Due to differences in institutional structures and sociocultural beliefs, most countries initiate and adjust NPIs dynamically in response to local epidemic transmission dynamics~\cite{rinaldi2022epidemiological,franks2022reopening,bin2021post}. We set three activation thresholds for NPIs under the aforementioned moderate initial scale of undetected transmission, such that they are activated when the current number of positive cases on the previous day exceeds 5\% (the low activation threshold), 15\% (the moderate activation threshold), and 35\% (the high activation threshold) of the urban population, respectively. NPIs are deactivated when the current number of positive cases falls below the corresponding threshold for two consecutive days.
	
	Compared to the baseline scenario, all NPIs reduce new infections and slow transmission during the intervention period (Figs. 2A–C). Specifically, lockdown and contact tracing achieve optimal containment across all activation thresholds, enabling one-time outbreak control and effectively preventing resurgences. Next are the 1 km activity radius restriction and the isolation strategy, while the closure of high-risk industries was the least effective. Moreover, aside from lockdown and contact tracing, all other NPIs lead to varying degrees of infection resurgence after their first implementation and subsequent lifting. The higher the activation threshold, the more severe the epidemic resurgence upon lifting under the observation period. Notably, the 1 km activity radius restriction and the isolation strategy show particularly sharp rebounds once lifted, whereas the closure of high-risk industries triggers a milder secondary resurgence. This disparity reveals a key mechanism: the intensity of the secondary rebound depends fundamentally on the size of the susceptible population in a city when the interventions are lifted. The 1 km activity radius restriction and the isolation strategy, due to their high control effectiveness during implementation, protect a large number of susceptible individuals from infection. However, once lifted, that susceptible population becomes the foundation for a major resurgence. Conversely, the less restrictive strategy (only shutting down high-risk industries) has already depleted a significant portion of susceptible individuals during the first wave, thereby limiting the potential for a rebound after lifting.
	
	At the economic level, as the intervention activation threshold increases, the rate of economic change during the first week shows a smaller decline, as shown in Figs. 2D-F. This is because, before the activation threshold is reached, uninfected individuals continue to engage freely in socioeconomic activities. However, as the pandemic progresses and NPIs are gradually implemented, the economic cost in later stages becomes increasingly severe. By restricting a large portion of the workforce from working, these measures reduce total social output while simultaneously incurring sustained public health and intervention cost. The combination of declining output and rising expenditure compounded, amplifying the economic cost over time. Notably, the isolation strategy incurs the highest economic cost among all NPIs, driven by the cost of PCR testing and isolation of pre-existing cases at implementation, as well as the recurring cost from epidemic fluctuations that trigger repeated activations. Moreover, under the high activation threshold, lockdown and contact tracing are the most effective in containing the epidemic, yet they incur greater economic cost than the baseline scenario. This is due to the high cost of implementing these strategies, including both the direct cost of isolation and the broader loss of workforce participation.
	
	We further examine the implementation impact of NPIs under the dynamic zero-COVID policy (0\%, the lowest activation threshold), as shown in Fig. 3. Dynamic zero-COVID policy is the primary epidemic control strategy adopted by China during the pandemic. It aims to rapidly detect and precisely contain local outbreaks while cutting transmission chains to prevent large-scale resurgences. Compared with flexible activation mechanisms, implementing NPIs under the dynamic zero-COVID policy yields clear advantages. Under this policy, all strategies achieve significantly lower infection rates and final infection scales while also greatly mitigating economic losses. Specifically, implementing lockdown and contact tracing achieve rapid elimination of cases, minimizing both new infections and the final outbreak size (Fig. 3A). Despite leading to high economic losses in the first week, both measures trigger a strong economic rebound once zero-COVID is achieved (Fig. 3B). According to our calculations, weekly economic cost under the two strategies recover to an average of 62\% and 16\% of pre-pandemic levels, respectively.
	
	\textbf{Outcomes of NPIs Under Different Initial Scales of Undetected Transmission.} To account for real-world delays in NPIs due to varying surveillance capabilities, we adopt a moderate activation threshold (i.e., the number of current positive cases reaching 15\% of the city's population) as the unified benchmark for intervention. Building on this, we examine the impact of NPIs under different initial scales of undetected transmission. Specifically, low and high initial scales correspond to 0.05\% and 3.5\% of the city's population, respectively, as the epidemic progresses naturally.
	
	Comparing NPIs under low (Figs. 4A-C) and high (Figs. 4D-F) initial scales of undetected transmission reveals a clear pattern: the larger the initial scale, the more positive cases accumulated in the early stage, and the sooner severe cases overwhelm healthcare capacity (Fig. 4D). This earlier overload, in turn, leads to earlier mortality events (Fig. 4E). Notably, under the same activation threshold, different initial scales of undetected transmission result in differences in the temporal distribution of pandemic dynamics and mortality events.
	
	As the initial scale of undetected transmission increases, the economic cost of all strategies occur earlier and become more severe overall, as shown in Fig. 4C and F. Moreover, larger initial scales of undetected transmission also amplify the dual health and economic fluctuations triggered by specific strategies after implementation. Specifically, for the 1 km activity radius restriction and the isolation strategy, which carry a higher risk of secondary resurgence, the rebound in severe cases is more pronounced, leading to greater fluctuations in economic losses.
	
	\textbf{Classification of NPIs’ Effectiveness.} We further develop a two-dimensional framework integrating containment effectiveness and economic cost to evaluate NPIs. To quantify how NPIs alleviate pressure on the healthcare system, we select three indicators for evaluation: the cumulative severe cases reduction rate, the peak severe cases inhibition rate, and the bed pressure alleviation rate (see Materials and Methods). To systematically measure the socioeconomic conditions during the interventions, we introduce the average weekly economic change ratio as a quantitative indicator (see Materials and Methods).
	
	As shown in Fig. 5, using the two indicators, we apply the DBSCAN clustering algorithm to group the strategies in a two-dimensional space defined by containment effectiveness and economic cost. Based on their characteristics, the five regional types (I-V) are defined as follows: low containment effectiveness-low economic cost (I), moderate containment effectiveness-low economic cost (II), high containment effectiveness-low economic cost (III), moderate containment effectiveness-high economic cost (IV), and high containment effectiveness-high economic cost (V). Our findings suggest that two types of strategies are not recommended for decision-making: Firstly, the type I includes strategies such as the closure of high-risk industries and the 1 km activity radius restriction under moderate-to-high activation thresholds. Although these strategies incur only moderate economic cost, their containment effectiveness remains extremely low (below 40\%). Notably, the closure of high-risk industries shows little sensitivity to either the initial scales of undetected transmission or activation thresholds, indicating limited practical value. Secondly, the type IV is represented by the isolation strategy activated at moderate-to-high thresholds. Following implementation, the large number of positive cases accumulated in the early phase leads to substantial expenditures and significant socioeconomic burdens.
	
	Types II and V receive relatively balanced overall evaluations. The former is exemplified by the 1 km activity radius restriction under the low activation threshold, which achieves moderate containment at relatively low economic cost. The latter include lockdown and contact tracing under the high activation threshold, and the isolation strategy under low (0\% and 5\%) activation thresholds. While these strategies achieve high containment effectiveness, they also incur relatively high economic cost. As such, they represent viable alternatives for countries or regions with different epidemic control priorities and resource constraints.
	
	Notably, we identify an optimal set of strategies—lockdown and contact tracing under low-to-moderate activation thresholds—located in Region III,with low (0\% and 5\%) activation thresholds performing particularly well. Such stringent strategies prove highly effective in containing the epidemic(exceeding 80\%) while minimizing economic cost (less than three times the pre‑epidemic level). These two strategies achieve a balance between high containment effectiveness and low economic cost. Moreover, both strategies maintain over 70\% containment effectiveness, regardless of the initial scale of undetected transmission or the activation threshold. However, their economic cost rises significantly with increasing activation thresholds. In contrast, the 1 km activity radius restriction, also located in Region III, shows a different trend: it is less sensitive to economic cost, but its containment effectiveness varies more strongly with the activation threshold, becoming increasingly dispersed across different threshold levels.
	
	\section{Discussion}
	The global pandemic of Coronavirus (COVID-19) has had a profound impact on international public health systems and the socioeconomic order, exposing systemic flaws in the existing global health governance framework. To prepare for future large-scale epidemics, the public health sector urgently needs to develop precision strategies for epidemic prevention and control. Our study introduces a two-dimensional framework for evaluating NPIs, grounded in a large dataset of fine-grained real-world individual mobility data from Shijiazhuang, and integrates two key dimensions: containment effectiveness and economic cost. By directly mapping NPI mechanisms onto individuals' daily behavioral patterns, it systematically evaluates the effectiveness and applicability of different NPIs.
	
	The implementation effects of NPIs vary across different activation mechanisms and initial scales of undetected transmission. Under a moderate initial scale of undetected transmission, lockdown and contact tracing achieve optimal containment outcomes across all activation thresholds. However, their economic cost increase progressively as the thresholds rise. This underscores the need for policymakers to carefully balance saving lives against economic cost. Among all strategies, the closure of high-risk industries is the least effective in containment, while the isolation strategy incurs the highest economic cost. As the activation threshold increases, strategies with relatively moderate containment effectiveness—such as the 1 km activity radius restriction and isolation—tend to activate more pronounced secondary epidemic rebounds. While raising the activation threshold mitigates economic losses in the short term, it leads to substantially greater losses later. Under the dynamic zero-COVID policy, lockdown and contact tracing strategies achieve the goal of eliminating infections within a short time. Although their economies declined during the implementation due to strict behavioral restrictions, a strong recovery followed once the strategies were lifted after achieving zero-COVID. As the initial scale of undetected transmission expands, the number of severe cases exceeds the healthcare capacity sooner, triggering healthcare system collapse earlier. Moreover, the mortality rate rises accordingly, and the peak in economic losses occurs significantly earlier. Unlike traditional models that often predict similar aggregate outcomes, our data‑driven approach reveals that these outcomes can exhibit different temporal distributions, which is critical for understanding the dynamics of healthcare burden and mortality.
	
	To systematically analyze the effectiveness of NPIs at different stages of epidemic, our study further constructs a two-dimensional evaluation framework based on both containment effectiveness and economic cost, thereby categorizing NPIs into five distinct types. We identify lockdown and contact tracing, particularly implemented under low activation thresholds (the dynamic zero-COVID policy and activation at 5\% of the urban population), as the optimal strategies. These two strategies achieve a balance between high containment effectiveness and low economic cost. The 1 km activity radius restriction strategy is highly sensitive to the activation threshold. By contrast, the closure of high-risk industries is not significantly affected by either the initial scales of undetected transmission or the activation threshold. It is worth noting that the isolation strategy invariably incur significant economic cost. While such measure can be relatively effective at controlling epidemics under low activation thresholds, policymakers must still weigh their cost against their benefits. Despite simulating only three initial scales of undetected transmission and activation thresholds, our two-dimensional framework clearly characterizes and localizes NPI’s effectiveness trends. Even without covering all variable levels, the precise identification of trend characteristics provides sufficient reference for policymakers to evaluate the effectiveness of various NPIs under different pandemic scenarios.
	
	Although our study focuses on COVID-19, the constructed two-dimensional evaluation framework for containment effectiveness and economic cost demonstrates strong universality. With the advancement of technology and the increasing accessibility of mobile trajectory data, this framework can be extended to the prevention and control research and policy formulation for other respiratory public health emergencies. It requires only the adjustment of model parameters based on the transmission characteristics and epidemiological patterns of different respiratory infectious diseases, thereby offering a reusable analytical tool and decision support for their precision containment.
	
	Our study also has certain limitations: Firstly, the NPIs we used are all based on idealized assumptions, such as a comprehensive public health management system and precise PCR testing capabilities. In practice, factors such as deviations in management implementation, constraints on test resources, and reporting delays may cause discrepancies between actual outcomes and model predictions. Secondly, our focus has been on evaluating the macro-level benefits of implementing NPIs at the city level without considering their micro-level impact on individual behavior. Future research could define such factors as individual tolerance levels and develop corresponding metrics to quantitatively evaluate public acceptance of activity restriction of varying intensity. Finally, our study primarily relies on data from a single city for its simulations, failing to adequately account for regional variations. The actual effectiveness of NPIs can be influenced by factors such as differences in economic development levels, healthcare resource allocation and population mobility patterns across cities.
	
	Despite these limitations, the evaluation framework developed in our study lays an important foundation for quantifying the effectiveness of NPIs. This framework, driven by real-world mobility data, enables a systematic evaluation of the comprehensive impact of NPIs across both epidemic containment and economic dimensions, providing quantifiable scientific evidence for public health decision-making. Against the backdrop of the World Health Organization's adoption of the Pandemic Agreement, we anticipate that this methodological framework will provide scientific tools for building a global epidemic prevention community, helping to strike the optimal balance between epidemic control and socioeconomic development.

	\section{Materials and Methods}
	\textbf{Data Sources and Preprocessing.}
	The raw data used in this study consist of a 7-day full-sample 4G communication record dataset from base stations and mobile terminals in Shijiazhuang, covering May 22-28, 2017, provided by one of the three major service providers. This dataset covers approximately 3 million anonymized users in Shijiazhuang, accounting for about one-third of the city's total population. In 2017, Shijiazhuang had a total of 11,594 base stations within its administrative boundaries. Over 7,500 of these base stations were located within the main urban area, covering approximately 100 square kilometers, with high spatial accuracy. When an anonymized user's mobile phone communicates with a base station via the 4G network, their location is recorded as the geographic coordinates of the nearest base station. Since some mobile applications continuously exchange data with backend servers, the frequency of location recording can reach up to once per second when 4G is active or indoor Wi-Fi is not in use. It is worth noting that the data capture movement patterns recorded before the implementation of epidemic prevention policies and before the public developed awareness of disease control measures. This data accurately reflects the true mobility patterns of populations under normal societal conditions, thereby providing a reliable baseline for evaluating NPIs.
	
	Given that co-location visits are critical for disease transmission, we first discretize the data following Lucchini et al.~\cite{lucchini2021living} to better capture dwell interactions. Specifically, each day is partitioned into 96 time windows of 15 minutes each, with a dwell time threshold of 10 minutes per window. If an anonymized user spends 10 minutes or more at a location within a 15-minute time window, whether continuously or discontinuously, they are considered to have stayed there during that window, and their data are recorded accordingly. Conversely, if an anonymized user does not remain at any location for more than 10 minutes within a 15-minute time window, they are considered to have had no location stay during that time. That is, the user is deemed to be in a moving state during that time window, and their data is recorded as “moving”. Two users are considered to have a co-location visit when they stay at the same location within the same time window. Second, to identify active users with valid movement trajectories, we apply strict data screening criteria requiring each anonymized user to meet all of the following conditions within their 7-day continuous trajectory: (1) a location of residence (see below for the definition of residence); (2) records of mobility in all 7 days; (3) a location outside home recorded in at least one time window per day; (4) at most 30 “moving” time windows among all the 96 time windows per day; (5) less than 24 time windows at residence among the 60 time windows per day during the daytime which we defined as the period between 6am to 9am. After screening, 702477 trajectories satisfy the above requirements.
	
	Finally, due to the periodic nature of individuals' daily movement trajectories and the difficulty of fully reflecting the regulatory effects of NPIs using raw 7-day trajectory data, we concatenate preprocessed continuous 7-day movement trajectories five times. This ultimately yields 35 days of 15-minute discrete trajectory data for 702,477 users in Shijiazhuang. For a more detailed description of the dataset's characteristics, the preprocessing workflow, and data quality validation, refer to~\cite{zhao2023high}.

	\textbf{Individual Attribute Construction.}
	Our study fully accounts for individual heterogeneity in disease transmission. As shown in Fig. 1A, we construct the following individual attributes based on mobility data and by integrating multi-source datasets: age, residential location, workplace location, and industry. Capturing these key dimensions of heterogeneity provides a foundation for accurately evaluating NPIs’ effectiveness.
	
	\textbf{Individual’s age.} To characterize differences in health trajectories among individuals post-infection, we use individual age to reflect transition probabilities between health states within the model. Specifically, we infer age by combining mobility capabilities derived from individual trajectory data with age distributions from census data, following the method of Zhao et al.~\cite{zhao2023high}.
	
	\textbf{Individual’s residence and workplace locations.} To reflect the heterogeneity in contact probabilities among individuals across different location categories, we infer residence and workplace locations from trajectory data. An individual's residence is defined as the location where the user spent the longest duration, at least 6 consecutive hours, between 8:00 pm and 7:00 am on weekdays. Similarly, each individual's workplace is defined as the location where the user spent the longest duration, with a minimum of 3 consecutive hours, between 7:00 am and 8:00 pm on weekdays. Additionally, for individuals who don’t spend more than 3 consecutive hours at any single location during daytime hours, their workplace cannot be determined. Such individuals were considered unemployed and labeled as “unemployed”.
	
	\textbf{Individual’s Industry.} To evaluate the economic impact of NPIs, we assign individuals to work industries using an allocation model based on the target city's occupational distribution data and company records. Specifically, for individuals whose workplace contains a single company, we assign that company's industry. For workplaces with multiple companies, we follow Barbour et al.~\cite{barbour2019planning}, using each company's number of insured employees as a weight. Using this weight, we then assign each individual an industry based on the target city's employment distribution across industries. If no company is present at the base station for an individual's work location, we assign an industry coarsely based on the workplace's POI category~\cite{zhao2023high}. Ultimately, the industries assigned to individuals exhibit strong statistical consistency with empirical data, achieving a Pearson correlation coefficient of 0.8505 $(p<0.001)$.
	
	\textbf{NPI Implementation Mechanism.}
	The five NPIs selected in this study fall into three broad categories based on their intervention mechanisms (Fig. 1D): (1) restrictions on individual movement distances (including lockdown and the 1 km activity radius restriction); (2) PCR testing and isolation (including isolation and contact tracing); and (3) closures of specific industries (i.e., the closure of high-risk industries). Given that the widespread deployment of epidemic control strategies has profoundly altered individual daily behavior, it is necessary to adjust existing trajectory data to reflect the mechanisms of different NPIs.The specific NPI implementation mechanisms and the corresponding individual trajectory correction processes are as follows:
	
	Restrictions on Individual Movement Distances. Primarily categorized as: lockdown and the 1 km activity radius restriction. The lockdown strategy mandates all residents to remain at home, undergo daily PCR testing, and be isolated along with their household members if tested positive. During lockdown, all individual trajectories for the corresponding days are modified to their residence. Isolated individuals are relocated to designated isolation points and removed from the transmission process. For the 1km activity radius restriction, individuals may only travel to workplaces or communities within 1km of their home during the policy period. Any location beyond 1km from home is modified back to the home location.
	
	PCR testing and Isolation. Primarily categorized as: isolation and contact tracing. In both, individuals undergo daily PCR testing during the implementation period. For isolation, positive cases and their household members are isolated together upon detection. For contact tracing, upon detection of a positive case, their co-location contacts from the previous day are traced, and then both the case and these contacts are isolated. Under both strategies, isolated individuals are relocated to designated sites and removed from the transmission process.
	
	Closures of specific industries, specifically high-risk industries. First, we select the industry categories involving individual‑to‑individual interactions from the 19 national economic sectors in the statistical yearbook. This yields in the following six sectors: wholesale and retail trade; transportation, warehousing, and postal services; accommodation and food services; education; health and social work; and culture, sports, and entertainment. Each location is assigned the industry with the highest employee count as its industry attribute. When high‑risk industries are closed, both employees of those industries and individuals whose trajectories pass through affected locations are treated as being at home.
	
	\textbf{Individual Micro-Level Communication Model.}
	To characterize the detailed infection process of disease transmission, we first define a co-location visit event as two individuals visiting the same location during the same time period. This forms the basis for constructing a data-driven micro-level individual transmission model. The core assumption of our transmission model is that if an individual involved in a co‑location visit has completed the incubation period, the other exposed individual faces a risk of infection. Fig. 1C illustrates the daily trajectories of three individuals who co-visited an entertainment venue between 21:00 and 21:45. If one or two individuals become infected and complete their incubation period during this time, the remaining individual at that location would face a certain probability of infection. To reflect the probability of contact between individuals at different locations, we define interactions occurring both between and within three types of locations: (1) residence, recording interactions between individuals and their household members within the home; (2) workplace, recording interactions between individuals and colleagues at their workplace; (3) community, recording interactions between individuals and others in the community, such as those arising from shopping at a mall.
	
	Fig. 1B shows the health status transition of an individual. If a susceptible ($S$) individual experiences a co-location event with an individual who has completed the incubation period at a given location, there is a certain probability that the susceptible individual will develop into an exposed ($E$) individual. After a 1.7-day incubation period, this individual will develop into either an asymptomatic patient ($I_{\text{as}}$) or a pre-symptomatic patient ($P_{\text{s}}$) based on their own condition. If the individual becomes an asymptomatic patient, they have a certain probability of making a full recovery. If the individual becomes a pre-symptomatic patient, after a 2-day symptom development period, they will progress to either a mild ($I_{\text{ms}}$) or severe ($I_{\text{ss}}$) patient. Mild patients have a certain probability of recovering fully or developing long-covid symptoms~\cite{lam2024persistence,oh2024incident,li2024short}. Severe patients, in turn, face a risk of death.The probability of state transitions for individuals is determined by their age.
	
	Moreover, during the COVID-19 pandemic, the sudden surge of patients in a short period inevitably places an overwhelming burden on urban healthcare systems. Without considering the construction of makeshift hospitals, the existing healthcare infrastructure in cities would struggle to meet patient demand. Therefore, we introduce a healthcare system collapse scenario, assuming that severe patients would be hospitalized and no longer participate in subsequent transmission. A healthcare system collapse is deemed to occur when the number of current severe patients in a city exceeds the available hospital beds. At this point, the probability of death for severe patients is calculated as follows:
	
	\begin{equation}
		d_c = d_0 \times \left(1 + \frac{S}{H} \ln \frac{S}{H}\right)^{\alpha}
	\end{equation}
	
	where, $d_0$ the baseline mortality probability for severe patients, $S$ is the number of current severe patients in the city, $H$ is the capacity of the city's healthcare system, $\alpha$ is a parameter with different values for different age groups: 0 for the 0-14 age group, a random number between [2, 3] for the 15–60 age group, and a random number between (3, 4] for the over-60 age group.
	
	\textbf{Economic Evaluation Model for Implementing NPI.}To evaluate the economic cost of NPIs, we develop a economic evaluation model to quantify their cost and impacts (Fig. 1E). This model incorporates both the GDP, government healthcare expenditure, and the direct cost incurred from implementing NPIs, aiming to systematically analyze the multidimensional economic cost associated with epidemic containment. The total economic for implementing NPIs is as follows:
	
	\begin{equation}
		E = E_{\text{IO}} - E_{\text{H}} - C_{\text{NPI}}
	\end{equation}
	
	where, $E_{\text{IO}}$ is the total GDP across all industries, i.e.,$E_{\text{IO}} = \sum_{j=1}^{N} E_{j_{\text{IO}}}$, with $N=19$ is the total number of industries. $E_{\text{H}}$ is the government healthcare expenditure, which primarily consist of medical expenses covered for mild cases ($E_{\text{ms}}$) and severe cases ($E_{\text{ss}}$). $C_{\text{NPI}}$ is the implementation costs for NPIs, mainly including PCR testing ($C_{\text{PCR}}$) and isolation ($C_{\text{Isolation}}$). Specifically, we calculate the economic value added across all industries by incorporating input‑output data and following the methodology of Haw et al. ~\cite{haw2022optimizing}. The total economic value added for industry $j$ over the entire time period $T$ is defined as follows:
	
	\begin{equation}
		E_{j_{\text{IO}}} = \sum_{\tau=0}^{T} x_{j\tau} e_{j\tau} y_j^* \left(1 - \sum_{i=1}^{N} a_{ij}\right) = \sum_{\tau=0}^{T} x_{j\tau} e_{j\tau} \left( y_j^* - y_j^* \sum_{i=1}^{N} a_{ij} \right), \quad i = 1,2,\dots,N
	\end{equation}
	
	where, $\tau$ is a single time period (set to 7 days in our study), $x_{j\tau}$ is the openness level of industry $j$ during period $\tau$ (taking values of 0 or 1), $e_{j\tau}$ is the average work efficiency of industry $j$ during period $\tau$, where an individual's work efficiency is determined by their health status, $y_j^*$ is the total output of industry $j$, and $a_{ij}$ is the proportion of industry $i$'s output allocated to industry $j$. Thus, $y_j^* \sum_{i=1}^{N} a_{ij}$ represents the value of intermediate products used by industry $j$.

	\textbf{NPI Evaluation indices.}
	To comprehensively evaluate the overall effectiveness of NPIs, we establish an evaluation framework encompassing two dimensions: containment effectiveness and economic cost.
	
	To measure the contribution of NPIs in alleviating healthcare overload, we develop the following evaluation index for containment effectiveness:
	
	\begin{equation}
		I_{\text{epidemic}} = \sqrt[3]{\text{CSCR} \times \text{PSCI} \times \text{BPA}}
		\label{eq:epidemic_index}
	\end{equation}
	
	$CSCR$ (Cumulative Severe Case Reduction Rate) is defined as $(C_{\text{baseline}} - C_{\text{NPI}}) / C_{\text{baseline}}$, where $C_{\text{baseline}}$ is the cumulative severe cases in the baseline scenario (i.e., without NPIs) and $C_{\text{NPI}}$ is the cumulative severe cases under NPI implementation. It reflects the overall effectiveness of NPIs in reducing the total burden of severe cases. $PSCI$ (Peak Severe Case Inhibition Rate) is defined as $(P_1 - P_2) / P_1$, where $P_1$ is the first peak of current severe cases and $P_2$ is the second peak of current severe cases. It reflects the ability of NPIs to suppress severe case peaks during epidemic fluctuations. $BPA$ (Bed Pressure Alleviation Rate) is defined as $1 - (T_c - T_0) / (T_{\text{obs}} - T_0)$, where $T_0$ is the NPI implementation time, $T_c$ is the time severe cases first fall below bed capacity, and $T_{\text{obs}}$ is the observation period. It reflects the extent to which NPIs alleviate healthcare system pressure.
	
	To quantify the economic cost of NPIs, we develop the following evaluation index:
	
	\begin{equation}
		I_{\text{economic}} = \frac{E_{\text{NPI}}}{E_{\text{ori}}} \times \frac{1}{w}
		\label{eq:economic_index}
	\end{equation}
	
	where, $E_{\text{NPI}}$ and $E_{\text{ori}}$ are the economic values with and without NPI implementation, respectively, and $w=5$ is the observation periods.
	
	\section*{Reference}
	\bibliographystyle{unsrt}
	\bibliography{main} 
	
	\clearpage
	\section*{Figures}
	\begin{figure}[h!]
		\centering
		\includegraphics[width=16 cm]{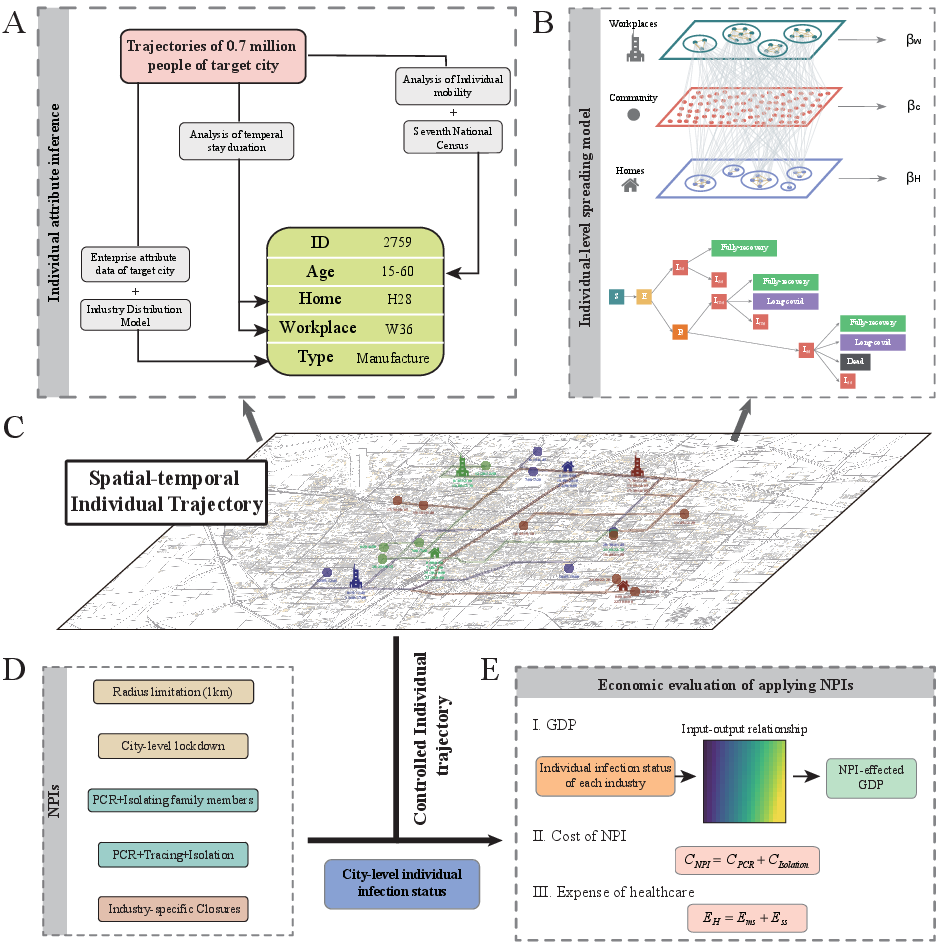}\\
		\caption{\textbf{Schematic diagram of the NPI evaluation methodology framework.}  (A) Individual attributes and their data sources. (B) A data-driven individual-level transmission model for cities. The upper panel shows disease transmission patterns within and between three types of locations, while the lower panel depicts the transition process of an individual's health status. (C) Schematic of individual spatiotemporal trajectories, illustrating a co-location event involving three individuals between 21:00-21:45 on a given day. (D) Five common types NPIs selected for our study. (E) Economic evaluation model for implementing NPIs.}\label{fig1}
	\end{figure}
	
	\begin{figure}[htbp]
		\centering
		\includegraphics[width=16 cm]{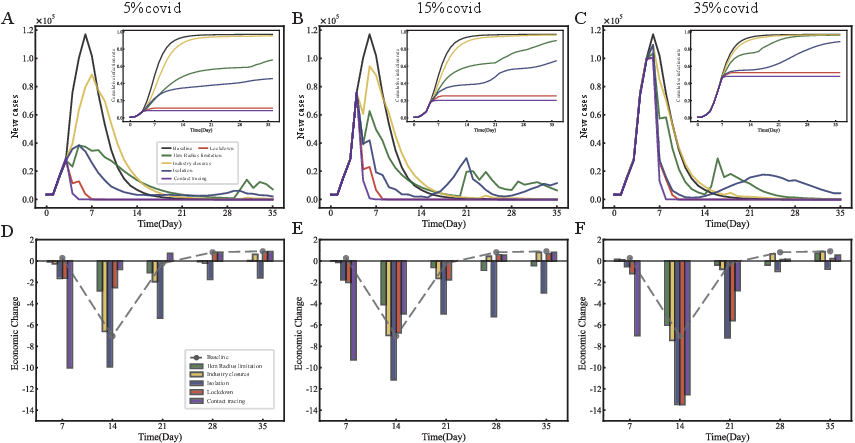}
		\caption{\textbf{Infections and economic cost under baseline scenarios versus NPIs under different activation thresholds.}Scenarios represent current positive cases reaching 5\%, 15\%, and 35\% of the city's population. (A)–(C) New cases (main figure) and cumulative infection rate (inset); (D)–(F) Weekly economic cost.}\label{fig2}
	\end{figure}
	
	\begin{figure}[htbp]
		\centering
		\includegraphics[width=16 cm]{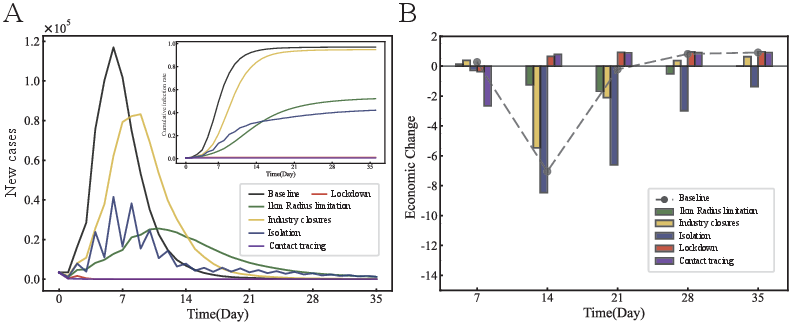}
		\caption{\textbf{Infections and economic cost under baseline scenarios versus NPIs under the dynamic zero-COVID policy.} (A) New cases (main figure) and cumulative infection rate (inset); (B) Weekly economic cost.}\label{fig3}
	\end{figure}
	
	\begin{figure}[htbp]
		\centering
		\includegraphics[width=16 cm]{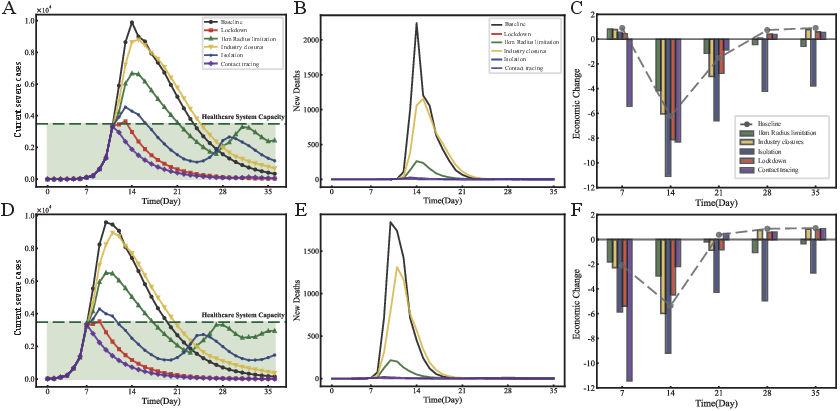}
		\caption{\textbf{Infections and economic cost under the baseline scenario versus NPIs implemented at a moderate activation threshold, for different initial scales of undetected transmission.} (A)(D) Current severe cases; (B)(E) New deaths; (C)(F) Weekly economic cost.}\label{fig4}
	\end{figure}
	
	\begin{figure}[htbp]
		\centering
		\includegraphics[width=16 cm]{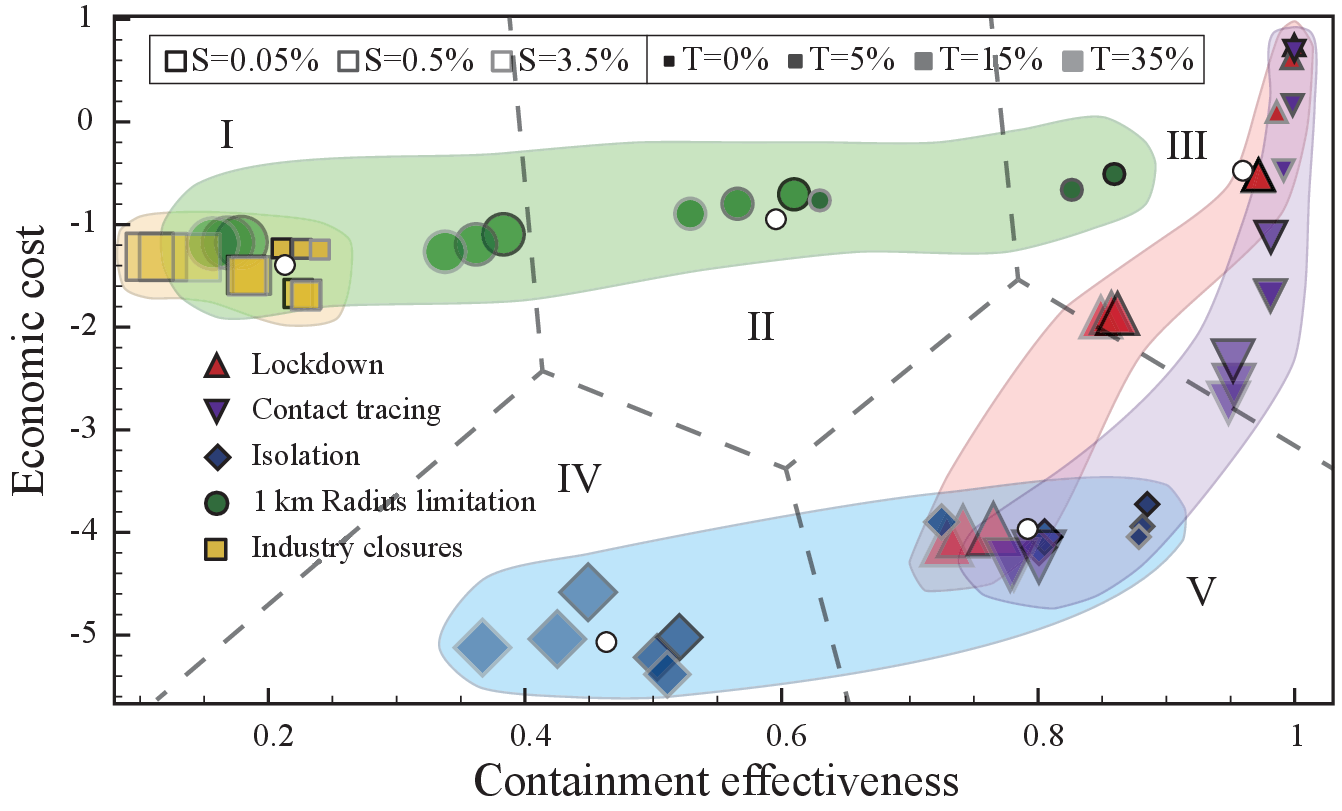}
		\caption{\textbf{Dual impacts of NPIs under varying initial scales of undetected transmission and activation thresholds.} Different shapes and colors represent distinct strategies. Shape size and color intensity indicate activation thresholds (0\%, 5\%, 15\%, and 35\%), where larger shapes and lighter colors correspond to higher thresholds. Border shading reflects initial scales of undetected transmission (0.05\%, 0.5\%, and 3.5\% of the city's population), with lighter borders indicating a larger initial scale of undetected transmission. Colored bands illustrate how the effectiveness of each NPI varies with the initial scale of undetected transmission and the activation threshold.}\label{fig5}
	\end{figure}

\end{document}